\documentclass[journal=jacsat,manuscript=article]{achemso}

\usepackage[version=3]{mhchem} % Formula subscripts using \ce{}

\author{Zhongyi Xia}
\affiliation[Strath]{Institute of Photonics, Department of Physics, University of Strathclyde, Glasgow, G1 1RD, Scotland, United Kingdom}
\altaffiliation{These authors contributed equally to this work}

\author{Dimitars Jevtics}
\affiliation[Strath]{Institute of Photonics, Department of Physics, University of Strathclyde, Glasgow, G1 1RD, Scotland, United Kingdom}
\altaffiliation{These authors contributed equally to this work}
\email{dimitars.jevtics@strath.ac.uk}

\author{Benoit Guilhabert}
\affiliation[Strath]{Institute of Photonics, Department of Physics, University of Strathclyde, Glasgow, G1 1RD, Scotland, United Kingdom}

\author{Jonathan J. D. McKendry}
\affiliation[Strath]{Institute of Photonics, Department of Physics, University of Strathclyde, Glasgow, G1 1RD, Scotland, United Kingdom}

\author{Qian Gao}
\affiliation[ANU]{{ARC Centre of Excellence for Transformative Meta-Optical Systems, Department of Electronic Materials Engineering, Research School of Physics, The Australian National University, Canberra, ACT 2600, Australian Capital Territory, Australia}}

\author{Hark Hoe Tan}
\affiliation[ANU]{{ARC Centre of Excellence for Transformative Meta-Optical Systems, Department of Electronic Materials Engineering, Research School of Physics, The Australian National University, Canberra, ACT 2600, Australian Capital Territory, Australia}}

\author{Chennupati Jagadish}
\affiliation[ANU]{{ARC Centre of Excellence for Transformative Meta-Optical Systems, Department of Electronic Materials Engineering, Research School of Physics, The Australian National University, Canberra, ACT 2600, Australian Capital Territory, Australia}}

\author{Martin D. Dawson}
\affiliation[Strath]{Institute of Photonics, Department of Physics, University of Strathclyde, Glasgow, G1 1RD, Scotland, United Kingdom}

\author{Michael J. Strain}
\affiliation[Strath]{Institute of Photonics, Department of Physics, University of Strathclyde, Glasgow, G1 1RD, Scotland, United Kingdom}
\email{michael.strain@strath.ac.uk}

\title[Modulation of nanowire emitter arrays using micro-LED technology]{Modulation of nanowire emitter arrays using micro-LED technology}

\abbreviations{IR,NMR,UV}

\begin{document}

%%%%%%%%%%%%%%%%%%%%%%%%%%%%%%%%%%%%%%%%%%%%%%%%%%%%%%%%%%%%%%%%%%%%%
%% The "tocentry" environment can be used to create an entry for the
%% graphical table of contents. It is given here as some journals
%% require that it is printed as part of the abstract page. It will
%% be automatically moved as appropriate.
%%%%%%%%%%%%%%%%%%%%%%%%%%%%%%%%%%%%%%%%%%%%%%%%%%%%%%%%%%%%%%%%%%%%%

%%%%%%%%%%%%%%%%%%%%%%%%%%%%%%%%%%%%%%%%%%%%%%%%%%%%%%%%%%%%%%%%%%%%%
%% The abstract environment will automatically gobble the contents
%% if an abstract is not used by the target journal.
%%%%%%%%%%%%%%%%%%%%%%%%%%%%%%%%%%%%%%%%%%%%%%%%%%%%%%%%%%%%%%%%%%%%%
\abstract{A scalable excitation platform for nanophotonic emitters using individually addressable micro-LED-on-CMOS arrays is demonstrated for the first time. Heterogeneous integration by transfer-printing of semiconductor nanowires was used for the deterministic assembly of the infrared emitters embedded in polymer optical waveguides with high yield and positional accuracy. Direct optical pumping of these emitters is demonstrated using micro-LED pixels as source, with optical modulation (on-off keying) measured up to 150 MHz. A micro-LED-on-CMOS array of pump sources were employed to demonstrate individual control of multiple waveguide coupled nanowire emitters in parallel, paving the way for future large scale photonic integrated circuit applications.}

%%%%%%%%%%%%%%%%%%%%%%%%%%%%%%%%%%%%%%%%%%%%%%%%%%%%%%%%%%%%%%%%%%%%%
%% Start the main part of the manuscript here.
%%%%%%%%%%%%%%%%%%%%%%%%%%%%%%%%%%%%%%%%%%%%%%%%%%%%%%%%%%%%%%%%%%%%%
\section{Introduction}\label{sec1}

Light generation and modulation at the nanoscale have attracted significant attention in recent years due to an availability of a broad range of advanced materials and the growing maturity of heterogeneous integration techniques \cite{Jevtics2022,Li2024,Wan2020}. Quasi-one-dimensional geometries, such as nanowires \cite{Jevtics2017,Yi2022,Yi2024,Kim2017,BermdezUrea2017,Wu2013,Takiguchi2017}, nanopillars \cite{DoloresCalzadilla2017}, nanobeam cavities \cite{Katsumi2019}, or cantilever-based emitters \cite{Chanana2022} are attractive solutions for on-chip emitters with efficient light coupling to waveguide platforms. The micro- or nano- dimensions of these emitters will play a crucial role in the next-generation of nanophotonics, due to their low emission threshold, high optical confinement, the ability to tune emission wavelengths, and importantly, the ability to directly integrate these emitters with other photonic technologies, such as waveguides, antennas, or grating structures \cite{Quan2019,Gniat2019}. 

The scalability of these devices from single device demonstrations towards on-chip systems requires advances in two key areas: (1) the controllable integration of multiple emitters with existing chipscale platforms and (2) the excitation of these devices using electronically controllable mechanisms that enable individual, high-speed addressing of multiple emitters in parallel. 

The fabrication and integration of nanowire emitters is typically based on the transfer of these devices from their growth substrate onto a receiver substrate, either with existing optical structures \cite{Jevtics2017, Xu2018} or where post-transfer fabrication processes are used to pattern waveguide circuitry around the transferred nanowires \cite{Yi2022,Yi2024}. The techniques of heterogeneous integration, such as micro-transfer-printing \cite{Guilhabert2016}, optical tweezers \cite{Pauzauskie2006}, micro-probe-based integration techniques \cite{BermdezUrea2017,Takiguchi2017}, are among the most popular for the integration of this geometry of devices into photonic systems \cite{Jevtics2022}, allowing for the use of large-scale device population pre-screening \cite{Jevtics2020} and the integration of devices into waveguides \cite{Takiguchi2017,Yi2022,Jevtics2017} to move towards engineered systems-on-a-chip.

The second challenge is that of scalable optical excitation of multiple nanowire sources in parallel, which is typically limited by the operation of continuous-wave or pulsed laser sources, which are generally used for their excitation. Laser pumps have a significant advantage over other means of excitation due to their coherent light emission, high peak power, and, where required, ultra-short pulses (reaching fs timescales) making them invaluable for fundamental device and materials studies. However, laser spot excitation in microscope pumping arrangements limits the optical addressing to single devices or clusters of photonic emitters at a time. Spatial light modulator (SLM) technology can be used to multiplex optical pump beams on a single substrate \cite{Panuski2022}, but the individual modulation of each sub-spot, beyond the common pulsed mode excitation, is limited to the switching bandwidth of the SLM device, which is in $\sim$ 10$^4$ Hz range for current digital mirror device systems \cite{Ren2015}.
Micro-LED-on-CMOS technology, on the other hand, has been rapidly advancing in functionality and scale over the last decade for applications in telecommunications \cite{Tsonev2014}, imaging \cite{Rae2009}, and displays \cite{Li2019}. Recent work from our group demonstrated a 128 $\times$ 128-pixel micro-LED-on-CMOS display with nanosecond pulsing rates, independent pixel control at frame rates up to 0.5 Mfps and 5-bit brightness control \cite{BaniHassan2022}.

\section{Results}

In this work we demonstrate the deterministic integration of multiple nanowire devices into waveguide arrays on-chip and the individual control of these nanowire emitters using an arrayed violet micro-LED-on-CMOS pump source \cite{McKendry2010}. By using electronically controllable LED array sources, we show individual addressing of multiple emitters in parallel and the route towards building scalable emitter systems for photonic integrated circuitry. The nanowire emitters used in this work are InP nanowires with a diameter of $\sim$ 660 nm and lengths in the range of 7 $\mu$m, see ref. \cite{Gao2014} for further information. The devices were fabricated using a bottom-up approach, with central emission wavelength at around 860 nm \cite{Guilhabert2016} and high-quantum efficiency, making them not only particularly bright, even at low excitation powers, but also ensuring their efficient absorption of the micro-LED light. These semiconductor emitters are robust and do not degrade during the transfer process, or under solvent treatment, which is advantageous for a multi-step fabrication process \cite{Jevtics2020, Yi2022}. 

 \begin{figure}[t!]
    \centering
    \includegraphics[width=1\textwidth]{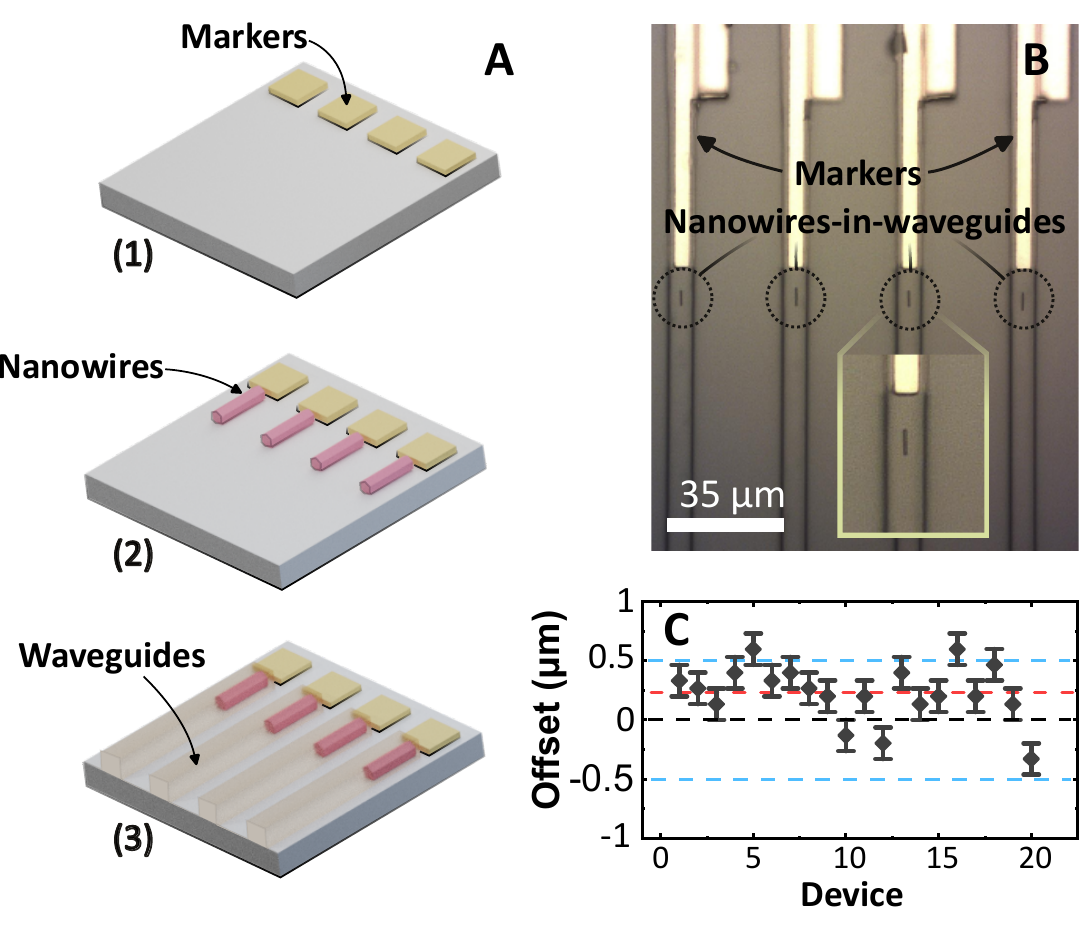}
    \caption{(\textbf{A}) Schematic flow diagram showing the fabrication process of embedding nanowire devices into polymeric waveguides. (\textbf{B}) Plan view image of the processed nanowire-in-waveguide arrays. Inset shows enlarged image of an embedded-into-waveguide nanowire. (\textbf{C}) A lateral offset scatter plot of the 20 nanowire-in-waveguide devices, with blue lines indicating $\pm$ 500 nm and red line showing the estimated average offset of 228 nm.}
    \label{fig1}
\end{figure}

After their growth, the nanowires were transferred to a host polymer (polydimethylsiloxane, PDMS) using a mechanical transfer process \cite{Jevtics2020}. The target nanowire devices were then individually selected for a further printing into the waveguides. A schematic of the fabrication process is shown in Fig. \ref{fig1}A, where a borosilicate glass was used as a substrate and metal marker structures were fabricated using direct write laser lithography and a metal lift-off process of a Ti:Au (50:200 nm) bilayer. A heterogeneous integration technique was then used to pick-and-place individual nanowires from the host PDMS substrate onto the glass substrate aligned to the central axis of the metal markers using previously developed alignment techniques \cite{McPhillimy2020}. After the nanowires were printed, a second laser lithography process, into a 4 $\mu$m-thick SU-8 polymer resist, was carried out to define the optical waveguide structures. This process resulted in nanowires fully embedded into the polymer waveguide, enhancing the optical mode coupling effciency between the structures \cite{Yi2022,Yi2024}, compared with end-fire or laterally coupled geometries \cite{Jevtics2017}. In total, 20 nanowire-in-waveguide devices were fabricated in a linear array, with waveguide spacing corresponding to the projected pitch of the elements of the micro-LED array (3:1 demagnification). A brighfield image of the embedded-nanowire array section is shown in Fig. \ref{fig1}B. The average offset of the nanowire major axis from the center of the waveguides was estimated to be 228 nm following the same estimation technique described in \cite{McPhillimy2020} and using a high-magnification optical microscope, as shown in the scatter plot in Fig. \ref{fig1}C.

\begin{figure}[h!]
 \centering
  \includegraphics[width=1\textwidth]{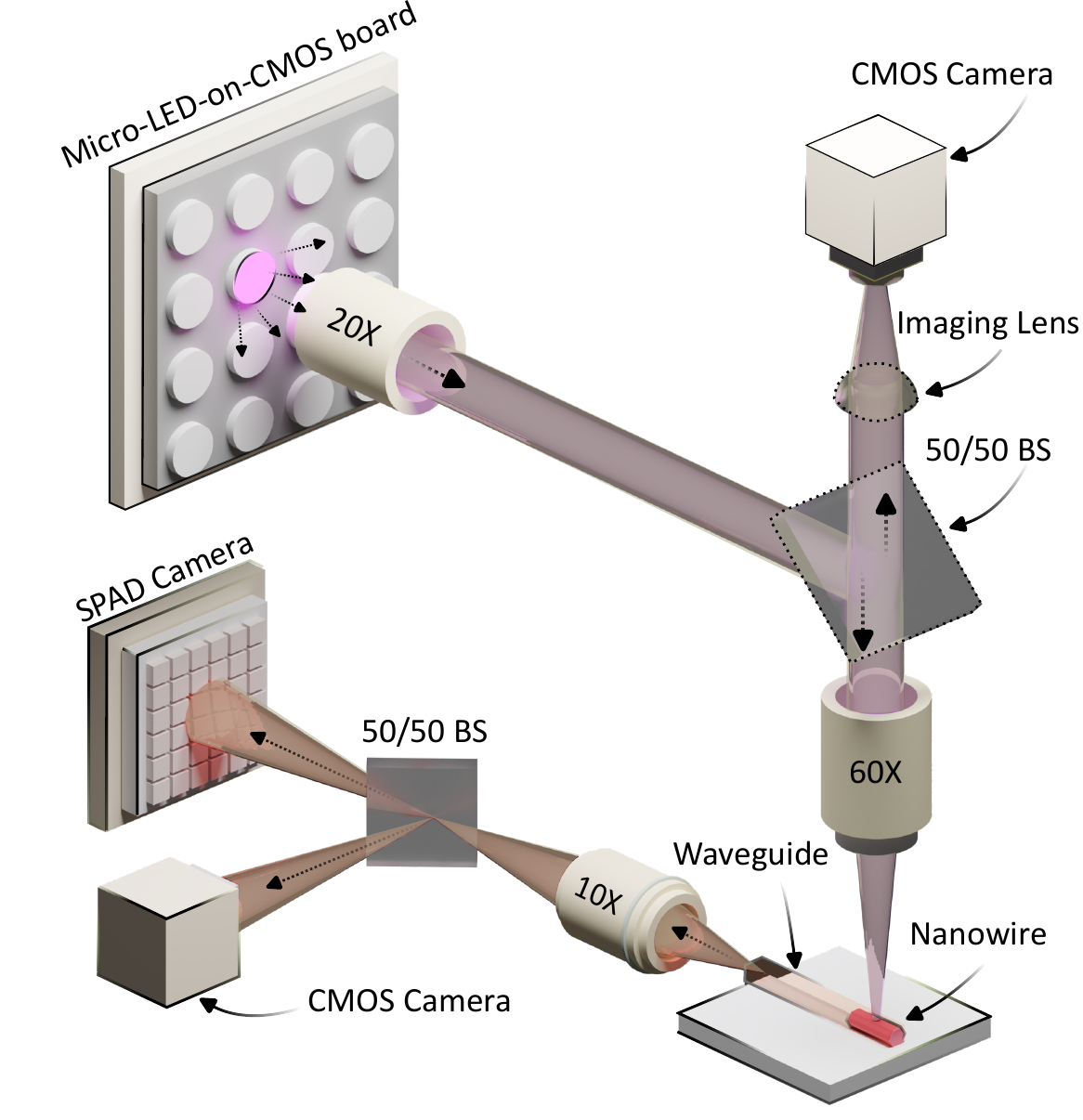}
 \caption{Schematic diagram showing the $\mu$-photoluminescence setup used for the optical excitation of nanowire emitters using micro-LED-on-CMOS technology.}
 \label{fig2}
 \end{figure}

The micro-LED-on-CMOS chip $\mu$-photoluminescence ($\mu$-PL) setup diagram for optical excitation and modulation of nanowire devices is shown in Fig. \ref{fig2}. The micro-LED-on-CMOS chip consists of a 16 $\times$ 16 active pixel array, where the diameter of the individual circular pixels is 72 $\mu$m, on a pitch of 100 $\mu$m. The optical power used here from an individual pixel in continuous-wave mode is of the order of 1 mW, with peak emission wavelength of 402 nm \cite{McKendry2010}. The cut-off frequency for the full on-off keying of the micro-LED pixels was measured at $\sim$ 120 MHz but this was limited here by the CMOS drive electronics since the micro-LEDs have bandwidth of up to 100's MHz \cite{McKendry2010}. The micro-LED-on-CMOS array is mounted on 3-axis stage allowing a fine alignment control with free-space optics. For the optical projection of micro-LED patterns onto a sample, a set of lenses are used to couple the projected light to the back-aperture of 60X optical objective. An optical bandpass filter was also used to filter out unwanted defect-related photoluminescence of GaN-based micro-LEDs \cite{Reshchikov2021}. A high quantum efficiency CMOS camera is used to image the emitted light from the micro-LED that is back-reflected through the optical column for alignment to the on-chip nanowire-in-waveguide devices. An edge detection setup is used to capture the light emitted from the end-facet of the optical waveguides, imaging through a 10X objective lens. The facet coupled light is then split and filtered using a 50:50 non-polarizing beamsplitter (BS) and bandpass filter into a high-sensitivity monochromatic CMOS camera and single-photon avalanche diode (SPAD) detector (PhotonForce32). The former is used for alignment and imaging purposes, the latter for time-domain light modulation measurements.

\begin{figure}[h!]
 \centering
  \includegraphics[width=1\textwidth]{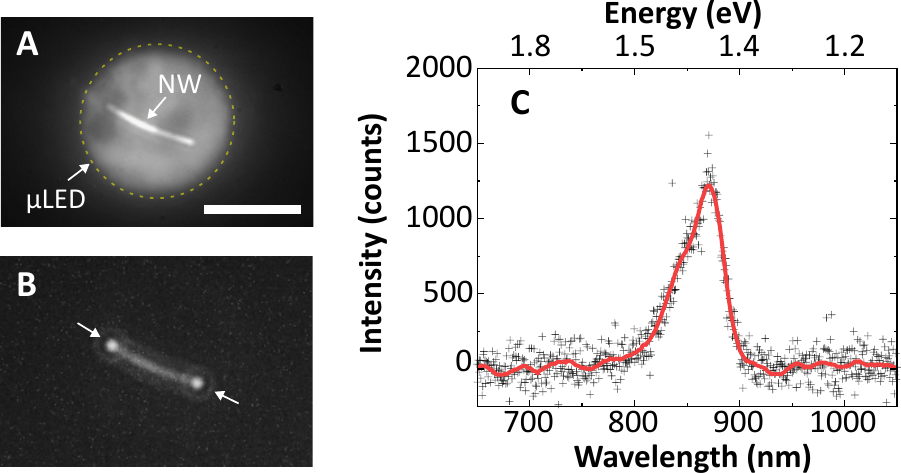}
 \caption{(A) Plan view optical image showing emission of a single micro-LED pixel projected onto a single nanowire emitter on quartz substrate. Scale bar = 12 $\mu$m. (B) Filtered image showing emission patterns coming out of the nanowire facets, excited with the projected micro-LED. (C) Spectrum from a single InP nanowire emitter on quartz under micro-LED optical excitation.}
 \label{fig3}
 \end{figure}

To characterize the nanowire emission properties, an array of nanowires was integrated onto a plain quartz disk substrate. The total micro-LED fluence (irradiance) of $\sim$ 280 mW/mm$^2$ was projected onto a sample, and an image of the micro-LED spot overlapping a single nanowire device is shown in Fig. \ref{fig3}A. From spatial overlap considerations an estimated $\sim$ 1$\%$ of the projected micro-LED illumination is incident on the single nanowire device.  The spatial mismatch of the micro-LED pixel and the emitter geometry could be improved by using shaped micro-LED apertures to match the profile of nanowire devices. Airy disks at each of the nanowire's facets are clearly visible on the pump-filtered micrograph in Fig. \ref{fig3}B. Furthermore, the emission wavelength of an optically excited nanowire, Fig. \ref{fig3}C, shows a measured nanowire emission spectrum taken under these low light micro-LED illumination conditions which is characteristic of the expected nanowire spontaneous emission \cite{Gao2014}.

%confirming the emission in the bandgap of 1.433 eV. Hence, we attribute this peak to the band edge emission from WZ InP NWs and the higher energy peak to the split off valence band.

\begin{figure}[h!]
 \centering
  \includegraphics[width=1\textwidth]{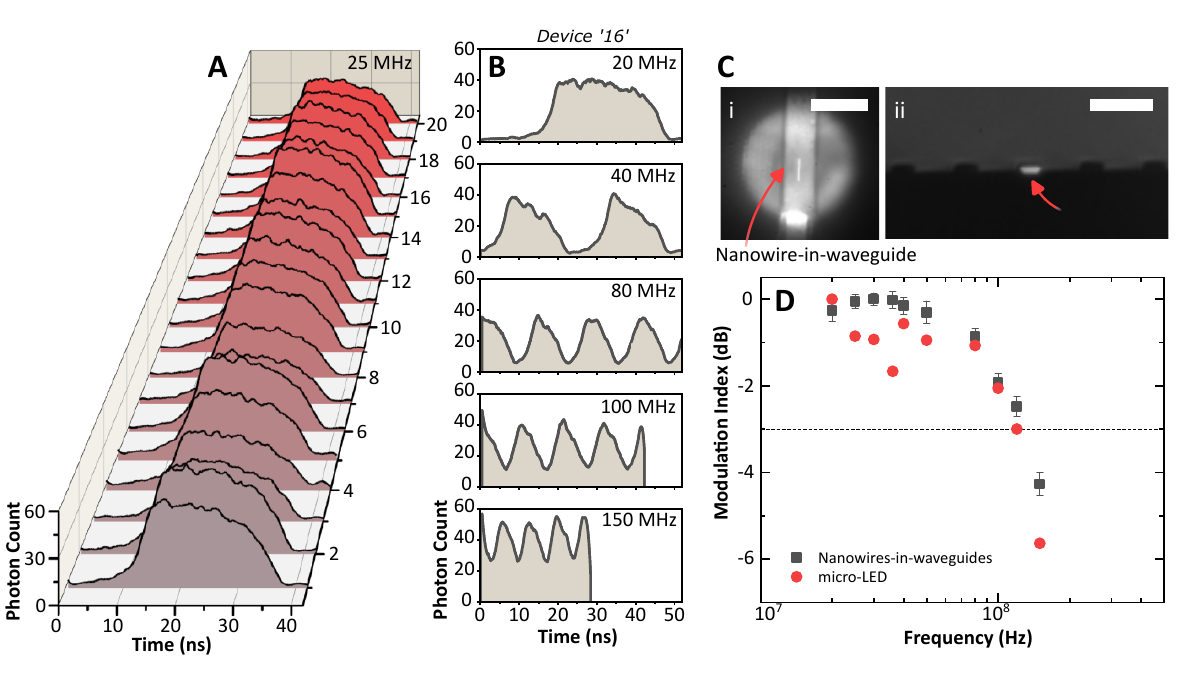}
 \caption{(A) Diagram showing time-domain measurements at modulation frequency of 25 MHz for 20 waveguide-integrated nanowire emitters. (B) Time-domain measurements of Device `16' at modulation frequencies of 20, 40, 80, 100, and 150 MHz, respectively. (C)(i) a brightfield micrograph showing the projected single pixel micro-LED emission exciting a single waveguide integrated nanowire and (ii) waveguide facet showing the light from the nanowire. Scale bars in (i) = 12 $\mu$m and in (ii) = 35 $\mu$m. (D) A cut-off frequency plot showing modulation index for both nanowire-in-waveguide emitters (mean value for 20 devices) and micro-LED pixel.}
 \label{fig4}
 \end{figure}

As described previously, target nanowire devices were embedded into polymeric SU-8 waveguides using a multi-stage micro-fabrication process. Each embedded nanowire emitter was optically excited and modulated (on-off keying) using the micro-LED-on-CMOS system. An external frequency source (Liquid Instruments, Moku:Pro) was connected to the CMOS board allowing to sweep between the excitation frequencies. Using the edge-detection setup, light collected at the facet of a nanowire-waveguide coupled device was captured using a SPAD camera. Due to the acquisition window of the SPAD camera, frequencies ranging from 20 - 150 MHz were selected, corresponding to 50 - 6.67 ns periods, respectively. To collect time-domain measurements of the modulated nanowire devices, both the micro-LED-on-CMOS board and a SPAD camera were connected to the outputs of an external frequency generator, for triggering the data collection; the frequency and phase of both outputs were synchronized. The  modulation characteristics of all 20 channels of the nanowire-in-waveguide devices were first measured independently using the SPAD camera. Fig. \ref{fig4}A shows a captured signal from all 20 nanowire-in-waveguide channels optically modulated at 25 MHz. The repeatability of the heterogeneous technique and integration into waveguides ensures that all 20 devices have comparable performance in terms of photon counts. The data was processed using Savitzky-Golay digital filter with a 3-rd order polynomial and a window size of 53 samples, for all modulation plots. Fig. \ref{fig4}B shows modulation measurements at different frequencies (20, 40, 80, 100, and 150 MHz) for a representative Device `16', demonstrating on-off keying operation beyond the the 3-dB cutoff frequency of the micro-LED devices. The top view microscope image in Fig. \ref{fig4}C(i) shows a micro-LED projected onto a waveguide, with a nanowire device located inside the waveguide structure. Side view of the edge of the sample showing output light at 20 MHz from the excited nanowire inside the waveguide is shown in Fig. \ref{fig4}C(ii). The measured cut-off frequencies, see Fig. \ref{fig4}D, of both micro-LED-on-CMOS board and embedded nanowire-in-waveguide devices (mean values) show that the latter follows the temporal envelope of the micro-LED emission, highlighting that the speed limitation originates with the pump as expected \cite{Takiguchi2017}. The modulation index is defined as log$_{10} \left( \frac{P_{max} - P_{min}}{P_{max} + P_{min}} \right )$, where $P_{max}$ and $P_{min}$ are maximum and minimum intensity values, respectively \cite{Takiguchi2017}. The nanowire devices exhibit slightly higher modulation index values as compared with the micro-LED devices. This is likely related to the amplified spontaneous emission operation of the nanowires which provides gain as a function of optical pump level, improving the signal to noise ratio. Furthermore, during the measurement of the low-light waveguide coupled experiments we observed luminescence from the SU-8 waveguide \cite{Pai2007}; however, it was characterized to be $\sim$ 28 times lower than the coupled emission from integrated-nanowire under the same optical pump conditions and using an 800 nm optical highpass filter. 

\begin{figure}[h!]
 \centering
  \includegraphics[width=1\textwidth]{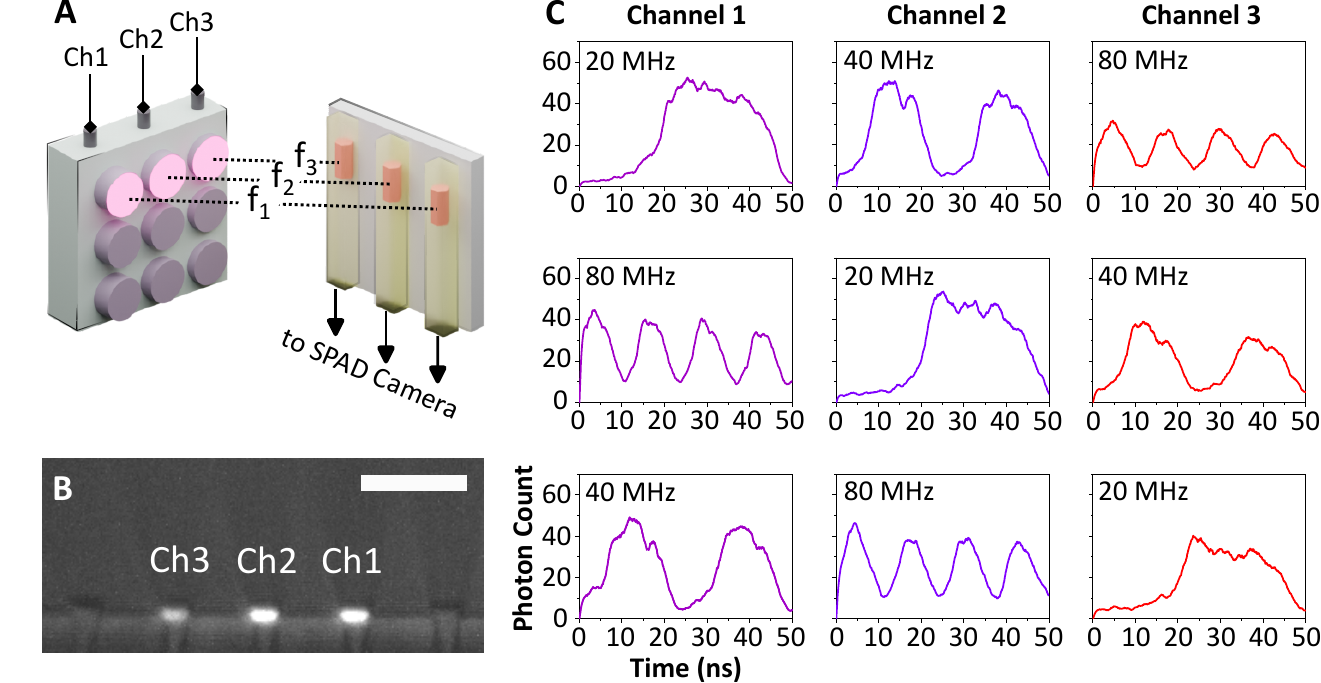}
 \caption{(\textbf{A}) A schematic diagram showing each of three adjacent and independently modulated micro-LED pixels separately exciting each of three adjacent embedded-into-waveguide nanowire emitters. (\textbf{B}) Side view of the edge of the waveguide sample showing output light from the three excited nanowires inside waveguides. Scale bar = 35 $\mu$m. (\textbf{C}) Time-domain modulation measurements (on-off keying) simultaneously collected from three waveguide embedded nanowires at frequencies of 80, 40, and 20 MHz, each horizontal panel representing a different configuration of modulation frequency.}
 \label{fig5}
 \end{figure}

To demonstrate the device excitation multiplexing capability, three adjacent micro-LEDs were projected onto three individual nanowire-in-waveguide devices at different frequencies (f$_1$, f$_2$, f$_3$), as shown schematically in Fig. \ref{fig5}A. By using free-space optical coupling from the micro-LED array, we can project three pixels without a significant power reduction of the excitation from the single pixel case. A different set of optics with a larger field of view, could significantly increase the number of projected micro-LEDs with a penalty on the optical throughput. The SPAD camera, consisting of a 32 $\times$ 32 array of pixels, and using 10X optical objective at the waveguide facet, allows imaging of the three waveguides in parallel. Each micro-LED pixel was independently modulated at 20, 40, and 80 MHz in different configurations. Fig. \ref{fig5}B shows three waveguide facets, each representing individual coupled nanowire emitters modulated at frequencies of 20, 40, and 80 MHz for channels 1, 2, and 3, respectively. Fig. \ref{fig5}C shows the results of independent frequency modulation of the three different channels of the sample with the individual waveguide channels measured on the SPAD array as isolated fields of view, in this case single pixels of the SPAD array. From photon count levels corresponding to each waveguide, it is clear that the waveguides have comparable signal levels, and the nanowire devices follow the modulated on-off keying of the micro-LEDs.

In summary, we have demonstrated the ability to create uniform arrays of nanowire-in-waveguide devices using a heterogeneous integration technique and post-transfer fabrication of polymer waveguides. The high yield of the device fabrication process is important for system scalability and enables the pre-selection of emitters with similar operating characteristics, or with advanced device binning, out of ensembles of nanowires. Optical pumping of nanowire emitters using micro-LED sources is demonstrated for the first time to our knowledge. The use of micro-LED-on-CMOS arrays as pump sources shows the capability for scalable, high-speed optical pumping of multiple emitters in parallel and experimental results show waveguide coupled optical signals in the 10’s of MHz range, limited in this case by the current CMOS driver chip.  The next stage of this work will be to integrate the optical pump system and nanowire-in-waveguide devices into a single package, creating a compact system and removing the need for projection optics. 

\subsection*{Acknowledgments}

This work was supported by the Royal Academy of Engineering (Research Chairs and Senior Research Fellowships), Engineering and Physical Sciences Research Council (EP/R03480X/1, EP/V004859/1, EP/W017067/1, EP/T00097X/1) and Innovate UK (50414). Z. Xia acknowledges PhD studentship support from Fraunhofer UK. The authors acknowledge the Australian Research Council for its financial support and Australian National Fabrication Facility, ACT node for providing to the NW samples. The authors also wish to thank H. Bookey from Fraunhofer UK and R. Henderson’s Group at the University of Edinburgh.

%%%%%%%%%%%%%%%%%%%%%%%%%%%%%%%%%%%%%%%%%%%%%%%%%%%%%%%%%%%%%%%%%%%%%
%% The appropriate \bibliography command should be placed here.
%% Notice that the class file automatically sets \bibliographystyle
%% and also names the section correctly.
%%%%%%%%%%%%%%%%%%%%%%%%%%%%%%%%%%%%%%%%%%%%%%%%%%%%%%%%%%%%%%%%%%%%%
\bibliography{acs-achemso}

\end{document}